\documentclass[12pt,showkeys,showpacs,indentfirst,bm]{revtex4}
\usepackage{epsfig}
\usepackage{epsf}
\usepackage{multirow}

\begin{document}

\title{Dynamically generated $0^+$ heavy mesons in a heavy chiral unitary approach}

\author{Feng-Kun Guo$^{1,2,6}$}
\email{guofk@mail.ihep.ac.cn}
\author{Peng-Nian Shen$^{2,1,4,5}$}
\author{Huan-Ching Chiang$^{3,1}$}
\author{Rong-Gang Ping$^{1,2}$}
\author{Bing-Song Zou$^{2,1,4,5}$}
\affiliation{\small $^1$Institute of High Energy Physics, Chinese
Academy of Sciences,
P.O.Box 918(4), Beijing 100049, China\footnote{Corresponding address.}\\
$^2$CCAST(World Lab.), P.O.Box 8730, Beijing 100080, China\\
$^3$South-west University, Chongqing 400715, China\\
$^4$Institute of Theoretical Physics, Chinese Academy of Sciences, P.O.Box 2735, China\\
$^5$Center of Theoretical Nuclear Physics, National Laboratory of
Heavy Ion Accelerator, Lanzhou 730000, China\\
$^6$Graduate University of Chinese Academy of Sciences, Beijing
100049, China}
\date{\today}

\begin{abstract}
In terms of the heavy chiral Lagrangian and the unitarized
coupled-channel scattering amplitude, interaction between the heavy
meson and the light pseudoscalar meson is studied. By looking for
the pole of scattering matrix on an appropriate Riemann sheet, a
$DK$ bound state $D_{s0}^*$ with the mass of $2.312\pm0.041$ GeV is
found. This state can be associated as the narrow $D_{sJ}^*(2317)$
state found recently. In the same way, a $B{\bar K}$ bound state
$B_{s0}^*$ is found, and its mass of $5.725\pm0.039$ GeV is
predicted. The spectra of $D_0^*$ and $B_0^*$ with $I=1/2$ are
further investigated. One broad and one narrow states are predicted
in both charm and bottom sectors. The coupling constants and decay
widths of the predicted states are also calculated.
\end{abstract}


\maketitle

\section{Introduction}

The recently discovered narrow-width state $D_{sJ}^*(2317)$
\cite{prl90} stimulates both experimental
\cite{prd68,prl91,prl92,ev04,ba04} and theoretical
\cite{go03,be03,cf03,dh03,fr04,nr04,ll04,cf05,gk05,mm05,ld06,%
ch03,cl04,ko05,nm05,te05,ww62,sz03,bc03,kl04,hl04,zc06,%
br03,ww63,bp04,vf06,bi05, wz05,ni05,ly06} interest. Many physicists
surmised that this new state is a conventional $c{\bar s}$ state
\cite{go03,be03,cf03,dh03,fr04,nr04,ll04,cf05,gk05,mm05,ld06}, and
the others believed that it can be an exotic meson state, such as a
four-quark state \cite{ch03,cl04,ko05,nm05,te05,ww62}, a $D_s\pi$
quasi-bound state \cite{sz03}, a $DK$ bound state
\cite{bc03,kl04,hl04,zc06}, a mixed state of $c{\bar s}$ with $DK$
\cite{br03,ww63} or with four-quark state \cite{bp04,vf06}, and etc.
On the other hand, one proposed that $D_{sJ}^*(2317)$ with $J^P=0^+$
could be the chiral partner of the ground state of $D_s$
\cite{be03,nr04}. However, the author in Ref. \cite{bi05} mentioned
that the chiral doubler produced by using Random Phase Approximation
equations should be ($D_s(1968)$,$D_s(2392)$) rather than
($D_s(1968)$,$D_s(2317)$), although the scalar state $D_s^*(2392)$,
as the scalar chiral partner of $D_s(1968)$ state, has not been
found yet \cite{bi05}. Up to now, the structure of $D_{sJ}^*(2317)$
is still indistinct and should carefully be studied. Moreover, the
Belle collaboration recently reported a broad $0^+$ charmed meson
with mass and width being $m_{D_0^{*0}}=2308\pm60$ MeV and
$\Gamma_{D_0^{*0}}=276\pm99$ MeV, respectively \cite{be04}, and the
FOCUS collaboration reported a broad $0^+$ charmed meson with mass
and width being $m_{D_0^{*0}}=2407\pm56$ MeV and
$\Gamma_{D_0^{*0}}=240\pm114$ MeV, respectively \cite{fo04}. Though
they are consistent with each other within experimental errors,
whether they are the same particle is still in dispute
\cite{te05,bl05}.

On the other hand, it has been shown that the light scalar mesons
$\sigma,~f_0(980),~a_0(980)$ and $\kappa$ can dynamically be
generated through the $S$ wave interaction between Goldstone bosons
in the chiral unitary approach (ChUA)
\cite{oo97,oo99,oond,ka98,na99,ma00,gp05}. In such an approach, the
amplitudes from the chiral perturbation theory (ChPT) are usually
adopted as the kernels of the factorized coupled-channel
Bethe-Salpeter (BS) equations. In this procedure, a Lagrangian in a
specific expanded order, where the symmetries of ChPT should be
preserved, is chosen at the beginning, and then the higher order
corrections to the amplitudes are re-summed with the symmetries kept
up to the order of the expansion considered. Namely, what the
unitary CHPT does in the successive step is re-summing a string of
infinite loop diagrams while the the symmetries of ChPT are held
\cite{ol00,ol01,ol03}. Moreover, ChUA has been applied to study the
$S$ wave interaction between the lower lying vector meson and the
Goldstone boson, and most of the known axial-vector mesons can also
be generated dynamically \cite{ro05}. Based on the valuable
achievements mentioned above, extending ChUA to the heavy-light
meson sector to study the $S$ wave interaction between the heavy
pseudoscalar meson and the Goldstone boson, and consequently the
structures of possible heavy scalar mesons, would be extremely
meaningful. In fact, similar work, called $\chi$-BS(3) approach, has
been done \cite{kl04,hl04}. In such an approach, heavy-light meson
resonances and open-charm meson resonances were predicted through
checking speed plots together with the real and imaginary parts of
the reduced scattering amplitudes. In our opinion, studying the
poles on the appropriate Riemann sheet of the scattering amplitude
would be a powerful procedure to reveal the properties of the
generated states in a more accurate way. In this paper, the $S$ wave
interaction between the heavy meson and the light pseudoscalar meson
is studied by using the extended chiral unitary approach, called
heavy chiral unitary approach. The poles that associate with the
experimentally observed narrow $D_{sJ}^*(2317)$ and broad $D_0^*$ in
the $I=0,~S=1$ and $I=\frac{1}{2},~S=0$ channels, where $I$ and $S$
denote the isospin and the strangeness, respectively, are searched.
The corresponding coupling constants and decay widths are also
discussed.


\section{Coupled-channel heavy chiral unitary approach}
\label{hcua}

In order to describe the interaction between the Goldstone boson and
the heavy pseudoscalar boson, we employ a leading order heavy chiral
Lagrangian \cite{bd92,wise,yc92}
\begin{equation}
\label{eq:L} {\cal L} =
\frac{1}{4f_{\pi}^2}(\partial^{\mu}P[\Phi,\partial_{\mu}\Phi]P^{\dag}
- P[\Phi,\partial_{\mu}\Phi]\partial^{\mu}P^{\dag}),
\end{equation}
where $f_{\pi}=92.4$ MeV is the pion decay constant, $P$
represents the charmed mesons $(c{\bar u},~c{\bar d},~c{\bar s})$,
namely $(D^0,~D^+,~D_s^+)$, and $\Phi$ denotes the octet Goldstone
bosons and can be written in the form of $3\times3$ matrix
\begin{equation}
\label{eq:ps} \Phi = \left(
\begin{array}{ccc}
\frac{1}{\sqrt{2}}\pi^0 + \frac{1}{\sqrt{6}}\eta & \pi^+ & K^+\\
\pi^- & -\frac{1}{\sqrt{2}}\pi^0 + \frac{1}{\sqrt{6}}\eta & K^0\\
K^- & \bar{K}^0 & - \frac{2}{\sqrt{6}}\eta
\end{array}
\right).
\end{equation}
This Lagrangian is equivalent to the $SU(4)$ extrapolation of the
ordinary meson meson chiral Lagrangian, eliminating the exchanges of
heavy vector mesons in the equivalent picture of vector meson
exchange \cite{hl05}. Obviously, the similar investigation in the
bottom sector can be carried out by replacing $P$ in Eq.
(\ref{eq:L}) with the anti-bottom mesons $(b{\bar u},~b{\bar
d},~b{\bar s})$, namely $(B^-,~{\bar B}^0,~{\bar B}_s)$.

We are interested in the heavy mesons in the $I=0,~S=1$ and
$I=\frac{1}{2},~S=0$ channels that can be specified by their own
isospins, respectively. In terms of Eq. (\ref{eq:L}), the amplitudes
can easily be obtained by
\begin{equation}
\label{eq:amp} V^{I}_{ij}(s,t,u) =
\frac{C^{I}_{ij}}{4f_{\pi}^2}(s-u),
\end{equation}
where $i$ and $j$ represent the initial state and the final state,
respectively. In the $I=0$ case, $i$ ($j$) can be 1 and 2 which
represent the coupled $DK$ and $D_s\eta$ channels in the charmed
sector, respectively, and $B{\bar K}$ and $B_s\eta$ channels in the
bottom sector, respectively. In the $I=\frac{1}{2}$ case, $i$ ($j$)
can take 1, 2 and 3 which denote the coupled $D\pi$, $D\eta$ and
$D_s{\bar K}$ channels in the charmed sector, respectively, and
$B\pi$, $B\eta$ and $B_sK$ channels in the bottom sector,
respectively. The coefficients $C^{I}_{ij}$ are listed in Table
\ref{tab:cij}.
\begin{table}[hbt]
\caption{\label{tab:cij} Coefficients $C^{I}_{ij}$ in
Eq.~(\ref{eq:amp}).}
\begin{center}
\begin{tabular}{ccc|cccccc}
\hline\hline $C^{0}_{11}$ & $C^{0}_{12}$ & $C^{0}_{22}$ &
$C^{1/2}_{11}$ & $C^{1/2}_{12}$
 & $C^{1/2}_{22}$ & $C^{1/2}_{13}$
 & $C^{1/2}_{23}$ & $C^{1/2}_{33}$\\
\hline $-2$ & $\sqrt{3}$ & 0 & $-2$ & 0 & 0 & $-\frac{\sqrt{6}}{2}$
& $-\frac{\sqrt{6}}{2}$ & $-1$\\
\hline\hline
\end{tabular}
\end{center}
\end{table}

The tree level amplitudes can be projected to the $S$ wave by using
\begin{equation}%
V^{I,~{\it {l}}=0}_{ij}(s)=\frac{1}{2}\int^{1}_{-1}d\cos{\theta}
V^I_{ij}(s,t(s,\cos{\theta}),u(s,\cos{\theta})),%
\end{equation}%
and
\begin{eqnarray}%
-u(s,\cos{\theta})&=& s-m_2^2-m_4^2 -
2\sqrt{[m_1^2+\frac{\lambda(s,m_1^2,m_2^2)}{4s}]
[m_3^2+\frac{\lambda(s,m_3^2,m_4^2)}{4s}]} \nonumber\\
&&
+\frac{1}{2s}\sqrt{\lambda(s,m_1^2,m_2^2)\lambda(s,m_3^2,m_4^2)}\cos{\theta},
\end{eqnarray}%
where $\lambda(s,m_i^2,m_j^2)=[s-(m_i+m_j)^2][s-(m_i-m_j)^2]$ and
the on-shell condition for the Mandelstam variables,
$s+t+u=\sum_{i=1}^{4}m_i^2$, is applied.

In ChUA, under the on-shell approximation, the full scattering
amplitude can be converted into an algebraic BS equation
\cite{oo97}
\begin{equation}
T=(1-VG)^{-1}V,
\end{equation}
where $V$ is a matrix whose elements are the $S$ wave projections of
the tree diagram amplitudes and $G$ is a diagonal matrix with the
element being a two-meson loop integral
\begin{equation}
G_{ii}(s)=i\int\frac{d^4q}{(2\pi)^4}\frac{1}{q^2-m_1^2+i
\varepsilon} \frac{1}{(p_1+p_2-q)^2-m_2^2+i\varepsilon},
\label{eq:2loop}
\end{equation}
where $p_1$ and $p_2$ are the four-momenta of the two initial
particles, respectively, and $m_1$ and $m_2$ are the masses of the
particles appearing in the loop. It was shown that the scattering
matrix derived in such a way satisfies the unitary relation
\cite{oo99,oond,gp05}.

The loop integral can usually be calculated in the center-of-mass
frame by using a three-momentum cut-off parameter $q_{max}$
\cite{oo97}. However, in this method, an artificial singularity of
the loop function might be produced \cite{gp05}, and the
applicability of the method is limited. The better way to remove the
singularity of the loop integral is using the dispersion relation
where a subtraction constant is employed. Then, the analytic
expression of $G_{ii}(s)$ can be expressed by \cite{oond}
\begin{eqnarray}%
G_{ii}(s)&=&\frac{1}{16\pi^2}\{a(\mu)+\log{\frac{m_1^2}{\mu^2}} +
\frac{\Delta-s}{2s}\log{\frac{m_1^2}{m_2^2}} \nonumber\\
&& +\frac{\sigma}{2s}[\log{(s-\Delta+\sigma)}
+ \log{(s+\Delta+\sigma)} \nonumber\\
&&- \log{(-s+\Delta+\sigma)} - \log{(-s-\Delta+\sigma)}] \},
\end{eqnarray}%
where $a(\mu)$ is the subtraction constant, $\mu$ denotes the
regularization scale,
$\sigma=[-(s-(m_1+m_2)^2)(s-(m_1-m_2)^2)]^{1/2}$ and
$\Delta=m_1^2-m_2^2$. This result is independent of $\mu$, because
the change in $G_{ii}$, caused by a variation of $\mu$, is cancelled
by the corresponding change of the subtraction constant $a(\mu)$.

\section{Poles on appropriate Riemann sheets} \label{pole}

The physical states are closely associated with the poles of the
scattering amplitude on the appropriate Riemann sheet of the energy
plane. For instance, considering only one channel, a bound state is
associated with a pole below the threshold value in the real axis of
the energy plane, and the three-momentum of the scattered meson in
the center of mass frame of the two mesons system can be written as
$p_{cm}=i|p_{cm}|$. A resonance should be related with a pole on the
second Riemann sheet, namely, Im$p_{cm}<0$. In the coupled channel
case, the situation is somewhat complicated. Detailed relation can
be found in Ref. \cite{bk82}.

Before searching for poles of the scattering amplitude, the range of
subtraction constant values in the dispersion relation method should
firstly be estimated. It can be done by comparing the calculated
value of loop integration in the dispersion relation method with the
one obtained in the cut-off method, although there might be an
artificial singularity problem in the cut-off method \cite{gp05}.
The cut-off momentum can approximately be chosen as
\begin{equation}
\label{eq:co} q_{max}\sim\sqrt{\Lambda_{\chi}^2-m_{\phi}^2},
\end{equation}
where $m_{\phi}$ is the mass of the Goldstone boson and
$\Lambda_{\chi}$ denotes the chiral symmetry breaking scale which is
about 1 GeV. The resultant $q_{max}$ for $\phi=\pi,~K$ and $\eta$
are all in the region of 0.8-0.9 GeV. Thus, it is reasonable to pick
up a value of $q_{max}$ in the region of $0.8\pm0.2$ GeV. Then, we
adjust the renormalization scale $\mu$ or the subtraction constant
$a(\mu)$ to match the calculated value of the loop integral in the
dispersion relation method with the one obtained in the cut-off
method at $\sqrt{s}=m_D(m_B)+m_K$ in a specific $q_{max}$ value
case, say $q_{max}=0.6$, 0.8 and 1.0 GeV, respectively. The
resultant loop integration curves versus $s$ in two different
methods are very close in the region around and below the matching
point $\sqrt{s}$. The corresponding values of $a(\mu)$ and $q_{max}$
are tabulated in Table \ref{tab:a}. With the estimated $a(\mu)$
value, the full scattering amplitude can be calculated.

\begin{table}[hbt]
\caption{\label{tab:a} The values of $a(\mu)$ from matching. We use
$\mu=m_{D}$ for the charm sector, and $\mu=m_{B}$ for the bottom
sector, respectively.}
\begin{center}
\begin{tabular}{cccc}
\hline\hline
$q_{max}$ (GeV) &~~ 0.6 &~~ 0.8 &~~ 1.0 \\
\hline
$a(m_D)$~~ &~~ -0.373~~ &~~ -0.630~~ &~~ -0.864 \\
$a(m_B)$~~ &~~~ 0.0232~ &~~ -0.0856~ &~~ -0.187 \\
\hline\hline
\end{tabular}
\end{center}
\end{table}

The poles of the scattering matrix in the $I=0,~S=1$ channel in both
the charmed sector and bottom sector are searched for first. It is
shown that on the first Riemann sheet of the energy plane, there is
only one pole located on the real axis below the lowest strong decay
threshold, $m_D+m_K=2.367$ GeV, in the charmed sector and only one
pole on the real axis below the lowest strong decay threshold,
$m_B+m_K=5.773$ GeV, in the bottom sector as well. The resultant
pole positions with different $a(\mu)$, which correspond to the
$q_{max}=0.6$, 0.8 and 1.0 GeV cases, are tabulated in Table
\ref{tab:i0}, respectively.
\begin{table}[hbt]
\caption{\label{tab:i0} Poles in the $(I,~S)=(0,~1)$ channel.}
\begin{center}
\begin{tabular}{cccc}
\hline\hline
$q_{max}$ (GeV)~~~~ &~~ 0.6 &~~ 0.8 &~~ 1.0 \\
\hline
$D_{s0}^*$ (GeV)~~~~  &~~ 2.353 ~~&~~ 2.317 ~~&~~ 2.270 \\
$B_{s0}^*$ (GeV)~~~~  &~~ 5.764 ~~&~~ 5.729 ~~&~~ 5.661 \\
\hline\hline
\end{tabular}
\end{center}
\end{table}
These poles are apparently associated with the $DK$ bound state and
the $B{\bar K}$ bound state, respectively. Due to the existence of
the ${\bar s}$ quark, these bound states should be scalar heavy
mesons, namely $D_{s0}^*$ and ${\bar B}_{s0}^*$, respectively. More
specifically, when $a(m_D)=-0.630$, corresponding to $q_{max}=0.8$
GeV, the mass of the $DK$ state, namely $D_{s0}^*$ , is about 2317
MeV, which is almost the same as the measured value of
$D_{sJ}^*(2317)$. Taking into account the uncertainty of subtraction
constant, the mass of the $D_{s0}^*$ (0, 1) state in our model is
$2.312\pm0.041$ GeV. Also due to the uncertainty of $a(m_B)$, the
predicted mass of the $B{\bar K}$ bound state, namely $B_{s0}^*$ (0,
1) state, is $5.725\pm0.039$ GeV. This mass is consistent with the
mass predicted in Refs. \cite{be03,bi05}, but larger than that in
Refs. \cite{mm05,kl04}. For comparison, we list the mass of
$B_{s0}^*$ predicted in different models in Table \ref{tab:bs0}.
\begin{table}[hbt]
\caption{\label{tab:bs0} Mass of $B_{s0}^*$ predicted in different
models.}
\begin{center}
\begin{tabular}{cccccc}
\hline\hline
  & Our result & \cite{be03} & \cite{bi05} & \cite{mm05} & \cite{kl04}\\
\hline ~~$m_{B_{s0}^*}$ (GeV)~~ &~~ $5.725\pm0.039$~~ &~~
$5.728\pm0.035$~~ &~~ $5.71\pm0.03$~~ & 5.627~~ &~~ 5.643\\
\hline\hline
\end{tabular}
\end{center}
\end{table}

In the $I=\frac{1}{2},~S=0$ case, the poles are located on
nonphysical Riemann sheets. Usually, if Im$p_{cm}$ is negative for
all the channels open for a certain energy, the width obtained would
correspond more closely with the physical one. We search for poles
in this particular sheet.

There are two poles in either charmed sector or bottom sector. The
width of the lower pole is broad and the width of the higher one is
narrow. The obtained poles are listed in Table \ref{tab:i1/2}.
\begin{table}[hbt]
\caption{\label{tab:i1/2} Poles in $(I,~S)=(\frac{1}{2},~0)$
channel.}
\begin{center}
\begin{tabular}{cccc}
\hline\hline
$q_{max}$ (GeV) & 0.6 & 0.8 & 1.0 \\
\hline
\multirow{2}{*}{$D_{0}^*$ (GeV)}  &~~ $2.115-i0.147$ ~~&~~ $2.099-i0.100$ ~~&~~ $2.079-i0.067$ \\
                                  &~~ $2.488-i0.039$ ~~&~~ $2.445-i0.049$ ~~&~~ $2.429-i0.002$ \\
\hline
\multirow{2}{*}{$B_{0}^*$ (GeV)}  &~~ $5.564-i0.160$ ~~&~~ $5.534-i0.110$ ~~&~~ $5.507-i0.074$ \\
                                  &~~ $5.864-i0.027$ ~~&~~ $5.827-i0.026$ ~~&~~ $5.821-i0.019$ \\
\hline\hline
\end{tabular}
\end{center}
\end{table}
In either the charmed or bottom sector, the lower pole is located on
the second Riemann sheet (Im$p_{cm1}<0$, Im$p_{cm2}>0$,
Im$p_{cm3}>0$, where $p_{cmi}$ denotes the momentum of one of the
interacting mesons in the $i$-th channel in the center of mass
system). This pole should be associated with a $D\pi$ ($B\pi$)
resonance in the charmed (bottom) sector. Consequently, this state
should easily decay into $D\pi$ ($B\pi$) in the charmed (bottom)
sector.

\begin{figure}[htb]
\begin{center}
{\epsfysize=3cm \epsffile{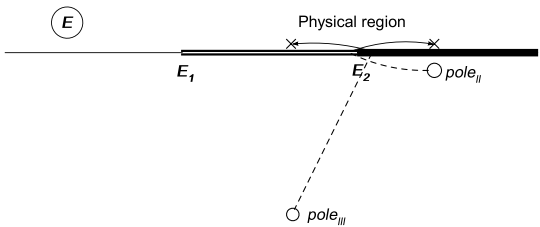}}%
\vglue -0.5cm\caption{\label{fig1}Paths from $pole_{II}$ on Riemann
sheet II and $pole_{III}$ on Riemann sheet III to the physical
region in the energy plane, where $E_1=m_{D}+m_{\pi}$ and
$E_2=m_{D}+m_{\eta}$.}
\end{center}
\end{figure}

The higher pole in either charmed or bottom sector is found on the
third Riemann sheet (Im$p_{cm1}<0$, Im$p_{cm2}<0$, Im$p_{cm3}>0$)
when $a(\mu)$ corresponds to $q_{max}=0.6$ GeV or 0.8 GeV, or on the
second Riemann sheet when $a(\mu)$ corresponds to $q_{max}=1.0$ GeV.
The pole should be associated with an unstable $D_s{\bar K}$
($B_sK$) bound state in the charmed (bottom) sector due to its
narrow width. It should be mentioned that the situation for the
higher pole in the later case, namely $a(\mu)$ corresponding to
$q_{max}=1.0$ GeV, is somewhat complicated. Besides a pole on the
second Riemann sheet, $pole_{II}=2.429-i0.002$ GeV shown in Table
\ref{tab:i1/2}, there is a shadow pole, $pole_{III}=2.397-i0.043$
GeV, on the third Riemann sheet. Note that
Re($pole_{II})>m_D+m_{\eta}$ and Re($pole_{III})<m_D+m_{\eta}$. A
sketch plot for the paths of these two poles to the physical region
in the energy plane is shown in Fig. \ref{fig1}. From this cartoon,
one sees that $pole_{II}$ corresponds more closely with the physical
one. Therefore, we choose $pole_{II}=2.429-i0.002$ GeV as the
result. Similar complexity appears at $pole_{III}=2.488-i0.039$ GeV
in Table \ref{tab:i1/2}, due to the existence of
$pole_{V}=2.048-i0.020$ GeV. With the same reason, we disregard
$pole_{V}$.

Considering the deviations of the data caused by the uncertainty of
$a(\mu)$ Table \ref{tab:i1/2}, we predict the mass and the width of
the broad $D_{0}^*$ ($\frac{1}{2}$, 0) state as $2.097\pm0.018$ GeV
and $0.213\pm0.080$ GeV, respectively, and the mass and the width of
the narrow $D_{0}^*$ ($\frac{1}{2}$, 0) state as $2.448\pm0.030$ GeV
and $0.051\pm0.047$ GeV, respectively. In the same way, we forecast
the mass and the width of the broad $B_{0}^*$ ($\frac{1}{2}$, 0)
state as $5.536\pm0.029$ GeV and $0.234\pm0.086$ GeV, respectively,
and the mass and the width of the narrow $B_{0}^*$ ($\frac{1}{2}$,
0) state as $5.842\pm0.022$ GeV and $0.035\pm0.019$ GeV,
respectively.

Recalling the predictions in Refs. \cite{kl04,hl04}, we noticed that
by checking the reduced scattering amplitude curves in the speed
plot, the authors in Ref. \cite{kl04} found a broad state with mass
of $2138$ MeV and a narrow states with mass of $2413$ MeV in the
charmed sector, and by further adjusting free parameters in the
next-to-leading order to reproduce the $D_{s0}^*(2317)$ state with
mass of $2317\pm3$ MeV and the $D_0^*$ state with mass of
$2308\pm60$ MeV and width of $276\pm99$ MeV given in Ref.
\cite{be04}, the authors in Ref. \cite{hl04} obtained a broad state
with mass of $2255$ MeV and width of about $360$ MeV and predicted a
very narrow state with mass of $2389$ MeV. In the same way, the
authors in Ref. \cite{kl04} further predicted a broad state with
mass of $5526$ MeV and a narrow states with mass of $5760$ MeV and
width of about $30$ MeV in the bottom sector. It seems that our
predicted $D_0^*$ ($\frac{1}{2}$,~0) states are consistent with
those in Ref. \cite{hl04}, although they still deviate from the
experimental data \cite{be04,fo04}. It should be mentioned that
because of the large uncertainty in the data analysis and existence
of the predicted higher narrow state just around the $D_2^*(2460)$
region, the present model could not be disregarded rudely.

\section{Coupling constants and decay widths}
\label{decay}

The decay properties of predicted states are studied by making the
Laurent expansion of the amplitude around the pole \cite{ol05}
\begin{equation}
T_{ij}=\frac{g_ig_j}{s-s_{pole}}+\gamma_0+\gamma_1(s-s_{pole})+\cdots,
\end{equation}
where $g_i$ and $g_j$ are coupling constants of the generated
state to the $i$-th and $j$-th channels. $g_ig_j$ can be obtained
by calculating the residue of the pole \cite{oond}
\begin{equation}
g_ig_j=\lim_{s\to s_{pole}}(s-s_{pole})T_{ij}.
\end{equation}

In the case where $a(\mu)$ corresponds to $q_{max}=0.8$ GeV, we
calculate the residues of the poles, and consequently the coupling
constants. The resultant coupling constants for the $D_{s0}^*$ and
$B_{s0}^*$ ($D_{0}^*$ and $B_{0}^*$) states are tabulated in Table
\ref{tab:ccs1} (\ref{tab:ccs0}). From these tables, one sees that
the coupling constants again are consistent with the results in the
pole analysis. In the $(0,~1)$ channel, the coupling of $D_{s0}^*$
($B_{s0}^*$) to the $D_s \eta$ ($B_s \eta$) channel is weaker than
that to the $DK$ ($BK$) channel. This is because the $D_{s0}^*$
($B_{s0}^*$) state is the $DK$ ($B{\bar K}$) bound state. In the
$(\frac{1}{2},~0)$ channel, the coupling of the lower broad $D_0^*$
($B_0^*$) state to the $D\pi$ ($B\pi$) channel is stronger than the
coupling of the higher narrow one to the $D\pi$ ($B\pi$) channel;
the coupling of the lower state to the $D\pi$ ($B\pi$) channel is
stronger than that to the $D_s\bar{K}$ ($B_sK$) channel and the
$D\eta$ ($B\eta$) channel, and the coupling of the higher state to
the $D_s\bar{K}$ ($B_sK$) channel is stronger than that to the
$D\eta$ ($B\eta$) channel and the $D\pi$ ($B\pi$) channel. These are
consistent with the pole analysis for the lower pole being a $D\pi$
($B\pi$) resonance and the higher pole being the unstable bound
state of $D_s{\bar K}$ ($B_sK$).

\begin{table}[hbt]
\caption{\label{tab:ccs1} Coupling constants of the generated
$D_{s0}^*$ and $B_{s0}^*$ states to relevant coupled channels. In
this case, $g_1$ and $g_2$ are real. All units are in GeV.}
\begin{center}
\begin{tabular}{cccc}
\hline\hline $~$ & Masses &~~~$|g_1|$~~~&~~~$|g_2|$\\
\hline $D_{s0}^*$ ~~~~&~~2.317~~&~~10.203~~&~~5.876  \\ \hline
$B_{s0}^*$ ~~~~&~~5.729~~&~~23.442~~&~~13.308  \\
\hline\hline
\end{tabular}
\end{center}
\end{table}
\begin{table}[hbt]
\caption{\label{tab:ccs0} Coupling constants of the generated
$D_{0}^*$ and $B_{0}^*$ states to relevant coupled channels. All
units are in GeV.}
\begin{center}
\begin{tabular}{cccccccc}
\hline\hline
 $~$ &  Poles  & $g_1$ & $|g_1|$ & $g_2$ & $|g_2|$ & $g_3$ & $|g_3|$\\
\hline $D_{0}^*$ & $2.099-i0.100$ & $7.750+i5.191$ & 9.328 & $-0.184+i0.096$%
                            & 0.208 & $4.648+i3.083$ & 5.578 \\
       $D_{0}^*$ & $2.445-i0.049$ & $0.030+i3.636$ & 3.636 & $-6.845-i2.248$%
                            & 7.205 & $-10.815+i1.543$ & 10.924 \\
\hline $B_{0}^*$ & $5.534-i0.110$ & $21.443+i12.060$ & 24.602 & $-2.239-i0.730$%
                            & 2.355 & $13.503+i7.016$ & 15.217 \\
       $B_{0}^*$ & $5.827-i0.026$ & $0.256+i6.958$ & 6.963 & $-14.697-i4.880$%
                            & 15.486 & $-25.000-i0.602$ & 25.003 \\
\hline\hline
\end{tabular}
\end{center}
\end{table}

The decay widths of generated states are further evaluated. We
first study the states in the $(0,~1)$ channel. The $D_{s0}^*$
state cannot decay into either $DK$ or $D_s\eta$, because the mass
of the state is lower than the threshold of the $DK$ channel.
Moreover, the $D_{s0}^{*+}(2317)\to D_s^+\pi^0$ decay violates the
isospin symmetry. Thus, the decay width of $D_{s0}^{*+}(2317)$
should be very small. This decay can only occur through
$\pi^0$-$\eta$ mixing. According to Dashen's theorem \cite{dash},
the $\pi^0$-$\eta$ transition matrix should be
\begin{equation}
t_{\pi\eta}=\langle\pi^0|{\cal H}|\eta\rangle=-0.003 ~\text{ GeV},
\end{equation}
and the decay width reads
\begin{equation}
\Gamma=\frac{p_{cm}}{8\pi
M^2}|\frac{g_2t_{\pi\eta}}{m_{\pi^0}^2-m_{\eta}^2}|^2,
\end{equation}
where $M$ is the mass of the initial state, $g_2$ represents the
coupling of $D_{s0}^*(2317)$ to $D\eta$, and $p_{cm}$ denotes the
three-momentum in the center of mass frame and can be written as
\begin{equation}
p_{cm} =
\frac{1}{2M}\sqrt{(M^2-(m_{D^+}+m_{\pi^0})^2)(M^2-(m_{D^+}-m_{\pi^0})^2)}.
\end{equation}
Then, the partial decay width of the $D_{s0}^{*+}(2317)\to
D_s^+\pi^0$ process can be obtained as
\begin{eqnarray}
\Gamma(D_{s0}^{*+}(2317)\to D_s^+\pi^0)&=& 8.69~\text{ keV}.
\end{eqnarray}
This value is compatible with that in Ref. \cite{ni05,ly06}.
Similarly, the partial decay width of the isospin violated decay
$B_{s0}^{*0}(5729)\to B_s^0\pi^0$ can be evaluated as
\begin{eqnarray}
\Gamma(B_{s0}^{*0}(5729)\to B_s^0\pi^0)&=& 7.92~\text{ keV}.
\end{eqnarray}

We then study the states in the $(\frac{1}{2},~0)$ channel. For
the higher state, two strong decay channels are opened. The
fraction ratio of the decay widths for these two decay channels
can be calculated by utilizing the coupling constants given in
Table \ref{tab:ccs0}. Let $\Gamma_1$ and $\Gamma_2$ denote the
partial decay widths with the final states being $D(B)\pi$ and
$D(B)\eta$, respectively. The ratio $\Gamma_1/(\Gamma_1+\Gamma_2)$
can be written by
\begin{equation}
R\equiv\frac{\Gamma_1}{\Gamma_1+\Gamma_2} =
\frac{|g_1|^2p_{cm1}}{|g_1|^2p_{cm1}+|g_2|^2p_{cm2}}.
\end{equation}
For higher $D_0^*$ and $B_0^*$ states, we have
\begin{equation}
R(D_0^*)=0.446,\qquad R(B_0^*)=0.829.
\end{equation}
It is shown that in the bottom sector, the higher narrow state is
easier to decay into $B\pi$ than into $B\eta$, but in the charmed
sector, the higher narrow state can decay into $D\pi$ and $D\eta$ in
almost the same weight.

\section{Conclusion}

Based on the heavy chiral unitary approach, the $S$ wave interaction
between the pseudoscalar heavy meson and the Goldstone boson is
studied. By calculating full scattering amplitudes via an algebraic
BS equation, the poles on some appropriate Riemann sheets are found.
These poles can be associated with bound states or resonances. With
a reasonably estimated single parameter $a(\mu)$ in the loop
integration, a pole on the real axis on the first Riemann sheet,
which is associated with the bound state, in the two-coupled-channel
calculation in the $(0,~1)$ channel is found. Because the mass of
the pole in the charmed sector is about $2.312\pm0.041$ GeV, this
state should be a $0^+$ $DK$ bound state and can be regarded as the
recently observed $D_{sJ}^*(2317)$. Meanwhile, a $0^+$ state
$B_{s0}^*$, which should be a $B{\bar K}$ bound state, is predicted.
Its mass is about $5.725\pm0.039$ GeV. In the $I=\frac{1}{2},~S=0$
case, three-coupled-channel calculations are performed in both
charmed and bottom sectors. In the charm sector, a broad pole
structure, which is associated with a resonance, is found at about
($2.097\pm0.018-i0.107\pm0.040$) GeV. Besides, a narrow pole
structure, which can be interpreted as a quasi-bound state of
$D_s{\bar K}$, at about ($2.448\pm0.030-i0.026\pm0.024$) GeV is also
found. In the bottom sector, one broad and one narrow poles are
found at about ($5.536\pm0.029-i0.117\pm0.043$) GeV and
($5.842\pm0.022-i0.018\pm0.010$) GeV, respectively. The coupling
constants of the generated states to the relevant coupled channels
are calculated. They are consistent with the results in the pole
structure analysis. In the $(0,~1)$ channel, the width of the
isospin violated decays $D_{s0}^{*+}(2317)\to D_s^+\pi^0$ and
$B_{s0}^{*0}(5729)\to B_s^0\pi^0$ are calculated. They are about
8.69 and 1.54 keV, respectively. Finally in the $(\frac{1}{2},~0)$
channel, the decay ratio $\Gamma_1/(\Gamma_1+\Gamma_2)$ for the
higher narrow state is also estimated.

\begin{acknowledgments}
We are very grateful to M.F.M. Lutz, H.-Y. Cheng, J.A. Oller and
Y.-L. Shen for valuable discussions. This work is partially
supported by the NSFC grant Nos. 90103020, 10475089, 10435080,
10447130, CAS Knowledge Innovation Key-Project grant No. KJCX2SWN02
and Key Knowledge Innovation Project of IHEP, CAS (U529).
\end{acknowledgments}


\end{document}